# Acousto-optic modulation of a photonic crystal nanocavity with Lamb waves in microwave K band

Semere A. Tadesse[1,2], Huan Li[1], Qiyu Liu[1], and Mo Li[1*]

[1]*Department of Electrical and Computer Engineering, University of Minnesota, Minneapolis, MN 55455, USA*

[2]*School of Physics and Astronomy, University of Minnesota, Minneapolis, MN 55455, USA*

*Corresponding author: moli@umn.edu*

ABSTRACT:

**Integrating nanoscale electromechanical transducers and nanophotonic devices potentially can enable new acousto-optic devices to reach unprecedented high frequencies and modulation efficiency. Here, we demonstrate acousto-optic modulation of a photonic crystal nanocavity using Lamb waves with frequency up to 19 GHz, reaching the microwave K band. The devices are fabricated in suspended aluminum nitride membrane. Excitation of acoustic waves is achieved with interdigital transducers with periods as small as 300 nm. Confining both acoustic wave and optical wave within the thickness of the membrane leads to improved acousto-optic modulation efficiency in the new devices than that obtained in previous surface acoustic wave devices. Our system demonstrates a novel scalable optomechanical platform where strong acousto-optic coupling between cavity-confined photons and high frequency traveling phonons can be explored.**

Exploitation of light-sound interactions in various types of media has led to a plethora of important optical technologies ranging from acousto-optic devices for optical communication[1] to photo-acoustic imaging in biomedicine[2]. Particularly, in the widely used acousto-optic devices, electromechanically excited acoustic waves in crystals are used to deflect, modulate and frequency-shift light to achieve optical signal processing[3]. Despite affording unique optical functionalities, conventional acousto-optic devices built on bulk crystals provide operation bandwidth only in the sub-gigahertz range, which is insufficient for the need of modern optical communication, and consequently has not been as widely employed as electro-optic devices[4]. With advances of integrated photonics and nanofabrication technology, it is now more feasible to miniaturize and integrate acousto-optic devices to augment their speed and performance so they can assume indispensable roles in integrated photonic systems for chip-scale optical communication[5]. Moreover, in addition to electromechanical excitation, acoustic waves or localized mechanical vibrations can also be optically stimulated through optomechanical forces including radiation pressure, gradient force and electrostriction. Such optomechanical effects recently has been extensively investigated in various optomechanical systems with dimensional scales ranging from meters to nanometers[6,7]. Therefore, with these recent developments, acousto-optics is entering a new era with plenty of research opportunities[8-11].

To enable strong and efficient acousto-optic interaction, one strategy is to simultaneously confine light and sound to optimize the spatial overlap of their modes. While three dimensional optical waveguides and cavities can be readily designed and fabricated, acoustic waves can be more conveniently excited electromechanically and confined in thin films which provide two dimensional confinement. Types of acoustic waves that are commonly employed for radio-frequency and microwave signal processing applications include bulk acoustic waves (BAW), surface acoustic waves (SAW), and in free-standing thin plates, Lamb waves (LW) and flexural plate waves (FPW)[12]. Among these, SAW and Lamb waves can be excited with periodically arranged planar electrodes named inter-digital transducers (IDTs), and thus are more suitable for integration with planar photonic systems[13,14]. The generated acoustic wave has a well-defined wavelength and frequency. The wavelength is determined by the IDT period ($\Lambda$), while the frequency depends on the wavelength and the sound velocity of the materials that constitute the structure in which acoustic wave propagates. Therefore, to achieve acoustic wave devices operating at ultrahigh frequency, the period of the IDT needs to be very small[15] and the materials with ultrahigh sound velocity, such as diamond[16], need to be used as the substrate.

To combine acoustic and photonic devices, the material platform needs to afford both piezoelectricity for excitation of acoustic waves and high refractive index contrast to enable optical confinement. Previously, we and other groups have used piezoelectric aluminum nitride (AlN) film deposited on silicon wafers with a layer of silicon dioxide ($SiO_2$)[11,17]. Since AlN has a relatively high refractive index of about 2.1, photonic waveguides and cavities can be fabricated in AlN with the $SiO_2$ layer as the cladding. At the same time, acoustic waves can be exited in the AlN layer. On this platform, we have demonstrated SAW wave with frequency up to 12 GHz and its acousto-optic modulation of optical ring resonators and photonic crystal nanocavities[11,18]. However, an important drawback of above devices based on SAW is that the $SiO_2$ layer has a lower sound velocity than both the top AlN layer and the bottom silicon substrate. As a result, the excited acoustic waves in the AlN layer tend to leak into and be guided in the $SiO_2$ layer whereas the optical modes are highly confined in the top AlN layer because of its high refractive index. Consequently, the modal overlap between the optical and acoustic modes is low, leading to relatively weak acousto-optic coupling efficiency. To circumvent this problem, in this work, we implement integrated acousto-optic devices on a suspended AlN membrane. When the membrane thickness is less than or comparable to the acoustic wave, the generated acoustic wave will propagate in Lamb wave mode with very high acoustic velocity. The removal of the substrate enforces the optical mode and the acoustic wave to maximally overlap within the thickness of the membrane. With this approach, we demonstrate acousto-optic modulation of a photonic crystal nanocavity at frequency up to 19 GHz with improved modulation efficiency.

The device consists of a photonic crystal nanobeam cavity and an IDT as shown in Fig. 1. The nanocavity is designed to support the fundamental dielectric mode to enhance the acousto-optic interaction in AlN[19]. The nanobeam is 800 nm wide and 46.8 μm long, inscribed with a nanocavity formed by an array of 600 nm wide rectangular holes arranged with a lattice constant of 520 nm. The length of the holes is varied adiabatically from 221 nm in the cavity center to 127 nm at the start of the mirror section of the nanocavity. An 800 nm wide waveguide is placed 1.0 μm away from the nanobeam to couple light into and out of the cavity. Input and output of light to/from the waveguide are through a pair of grating couplers. Two sets of IDTs with periods of 2.0 and 0.3 μm (corresponding to electrode linewidth of 500 nm and 75 nm), with 150 and 250 pairs of electrode fingers, respectively, are fabricated. The IDTs are placed to launch acoustic waves propagating in the direction transverse to the nanobeam and their apertures are designed to be the same as the nanocavity length of 46.8 μm to maximize acousto-optic overlap.

The devices were fabricated on an AlN (330 nm)/Si substrate. Fig. 1(a) and (b) show optical microscope images of a device with 2.0 μm period IDT. The photonic structure was first patterned in the AlN layer using electron beam lithography (EBL) and plasma etching processes. The 330 nm thick AlN layer was etched by 200 nm in depth, leaving a 130 nm slab to facilitate acoustic wave propagation and reduce acoustic reflection. Subsequently, the IDT electrodes were patterned using EBL and electron beam evaporation of 50 nm thick aluminum followed by a liftoff process. Another step of EBL and plasma etching was done to open windows (dark rectangular holes in Fig. 1(a)) for releasing the membrane. Finally, a $XeF_2$ dry etching process was used to release the membrane. A cross-sectional view illustrating the structure of the IDT, the waveguide and the nanocavity on the suspended membrane is shown in Fig. 1(c).

The nanocavity side-coupled to the waveguide was characterized by measuring its transmission spectrum using a tunable laser. The result is shown in Fig. 1(d), displaying several resonance modes of the nanocavity. We focus on the fundamental resonance mode (a symmetric mode) at 1543.94 nm with a loaded quality factor of $6.3 \times 10^3$ (corresponding to linewidth of 31 GHz) and an extinction ratio of over 25 dB (inset, Fig. 1(d)). The corresponding intrinsic quality factor is $1.1 \times 10^5$. The IDT transducers were characterized by measuring the spectrum of reflection coefficient $S_{11}$ using a vector network analyzer (VNA, Agilent E8362B). The VNA Port 1 was directly connected to the IDT through a microwave probe. As shown in Fig. 2(a) and (b), for the device with 2.0 μm period IDT, four prominent acoustic modes with frequencies of 1.35 GHz, 5.40 GHz, 10.12 GHz, and 14.46 GHz, respectively, can be observed; for the device with 0.3 μm period IDT, two modes with frequencies of 16.37 GHz and 19.20 GHz, respectively, can be observed. The resonance frequencies agree well with the eigenfrequencies calculated for a 330 nm thick AlN membrane using finite element method (FEM) simulation package[20]. The calculation yielded two fundamental Lamb wave modes (A0 and S0) and many higher order modes (A1, S1, A2, S2, etc.). The nomenclature of the modes is based on the symmetry of the displacement field profile and the number of nodes, with respect to a plane dissecting and in the plane of the membrane. The simulated displacement profile and strain field distribution of each mode are shown in Fig. 2(c). Within the measurable frequency range limited by the maximum frequency (20 GHz) of our VNA, the A0, S0, A1 and S1 modes can be observed in the device with 2.0 μm period IDT, whereas only the A0 and S0 modes can be observed in the device with 0.3 μm period IDT. Compared with the SAW wave generated with the same 2.0 μm period IDT but on unsuspended AlN layer on the $SiO_2$ (3 μm)/Si substrate, the frequency of Lamb wave on suspended membrane is almost an order of magnitude higher because the acoustic wave only propagates in AlN which has a high sound velocity. In addition, because the displacement and strain fields (hence the mechanical energy) of the Lamb wave are confined in the membrane that is completely decoupled from the substrate, the propagation loss is expected to be much less than SAW waves which has high loss to the substrate.

To study the frequency-wavelength dispersion relation of the Lamb wave in the AlN membrane, IDTs with a wide range of periods were fabricated, measured and simulated, as shown by the results in Fig. 2(d). Excellent agreement was obtained between experiment and theory. It can be observed that, for IDT periods much longer than the AlN film thickness (330 nm), the S0 mode has a much higher frequency (more than double) than that of the A0 mode. But as the IDT period deceases to be comparable or less than the thickness of the membrane, the A0 and S0 mode frequencies both approach to a value expected for a Rayleigh-type surface wave. Therefore, to support Lamb wave of very high frequency, the membrane thickness can be further reduced and the S0 Lamb mode has the advantage of higher frequency than the A0 mode.

To characterize the acousto-optic interaction between the Lamb wave and the optical mode of the nanocavity, the probe laser was blue-detuned to the nanocavity's fundamental resonance mode. The output optical signal was amplified with an erbium doped fiber amplifier (EDFA) and then filtered with a tunable optical filter to suppress amplified spontaneous emission noise introduced by the EDFA. The amplified and filtered optical signal was then sent to a high speed photoreceiver and the output electrical signal was returned to the VNA Port 2. The VNA Port 1 remained connected to the IDT. In this measurement configuration, the response of the nanocavity resonance to the acousto-optic modulation induced by the Lamb wave was characterized by measuring the spectrum of the system's transmission coefficient $S_{21}$, which was obtained by sweeping the frequency of the VNA output signal from Port 1 to the IDT. During the measurement, the VNA output power was fixed at 10 µW (-20 dBm) to minimize heating of the membrane, which can cause frequency shift of the nanocavity's resonance. The measured transmission spectrum for the two devices are shown in Fig. 3. The $S_{11}$ spectra are also plotted for comparison. As can be seen from the plots, the acousto-optic response spectra show peaks at frequencies matching those of the Lamb wave modes measured in the $S_{11}$ spectra. Comparing the amplitudes of the peaks reveals the different acousto-optic modulation strength of the various acoustic modes. Specifically, for the device with 2.0 µm period IDT, it can be observed that the higher order modes A1 and S1 induce much weaker (almost 30 times weaker) modulation than the fundamental modes A0 and S0. For the device with 0.3 µm period IDT, only the modulation of A0 and S0 modes was measured within the 20 GHz bandwidth of the VNA.

The acoustic wave modulates the nanocavity's resonance mode mainly through elasto-optic effect, in which the strain field of the acoustic wave dynamically changes the material's refractive index. Therefore, the variation of the acousto-optic modulation strength among the various Lamb wave modes is related to the different modal overlap between their strain field distribution and the electric field distribution of the nanocavity's resonance mode. More detailed theory of the acousto-optic modal overlap can be found in our previous work[11]. In addition, the varying excitation efficiency and propagation loss of the different acoustic modes also play a role in the overall efficiency. For example, as shown in Fig. 2(c), the A-modes have a strain field distribution that is antisymmetric with respect to a lateral plane dissecting the membrane, resulting in cancellation of elasto-optic effect when its overlap with the optical mode integrated over the thickness of the membrane. This is in contrast to the S-modes where the symmetric strain field distribution results in enhanced elasto-optic overlap integral with the optical mode, and thus enhanced acousto-optic modulation. On top of that, the S-modes have relatively higher excitation efficiency by the IDT than the A-modes, as can be seen from the $S_{11}$ spectra (S-modes have deeper dips than A-modes). This trend, however, is counteracted by the fact that the S-modes have higher frequencies than the A-modes and hence higher propagation loss. The dramatically reduced modulation strength of the higher order modes (A1 and S1) is attributed to their high propagation loss, since propagation loss of acoustic waves scales approximately as the square of the frequency[16]. Among Lamb waves of different wavelengths (0.3 and 2.0 µm), the modulation strength is highly dependent on the ratio of the acoustic wavelength to the width of the nanocavity, which is fixed at 0.8 µm in our devices.

Acousto-optical modal overlap is non-vanishing if the acoustic wavelength is more than twice the nanocavity width, which is the case for the device with 2.0 μm period IDT. For the device with 0.3 μm period IDT, the acousto-optic modulation is much weaker due to the vanishing modal overlap for it is integrated over the nanocavity's width of several acoustic wavelengths, as shown in Fig. 2(c). In addition to the main peaks corresponding to the well-defined acoustic modes, in the optical $S_{21}$ spectra shown in Fig. 3, there are several unidentified peaks and many ripples. We currently attribute them to the reflection of the acoustic wave at the complicated boundary of the suspended membrane (see Fig. 1(a)) and the interference it has caused. By introducing efficient absorbing structures around the devices, undesired reflection of acoustic waves can be reduced and those spurious peaks can be eliminated.

We quantify the devices' acousto-optic modulation efficiency ($G$) by performing the calibrated modulation measurement. The efficiency is defined as the change in the nanocavity's resonance frequency per square root of the microwave power sent to the IDT. The results are summarized in Table I. For the device with 2.0 μm period IDT, the fundamental modes A0 and S0 have an order of magnitude higher efficiency than the higher order modes A1 and S1. The S0 mode, with its combined high efficiency and high frequency, thus is more ideal for cavity optomechanical experiments. Although much weaker than its fundamental counterpart, the S1 mode is also interesting for its ultra-high frequency. For the device with 0.3 μm period IDT, the modulation efficiency for both A0 and S0 modes is much lower than that of the 2.0 μm period IDT. This is expected as the Lamb wavelength is a small fraction (almost a third) of the nanocavity's width, leading to a very small acousto-optic modal overlap integral. This problem can be solved by designing a nanocavity with a very small effective mode width, such as the dumbbell slot cavity design[21].

In summary, we have demonstrated a novel optomechanical system on a suspended AlN membrane on which very high frequency acoustic transducers and photonic crystal nanocavities are integrated. The system overcomes the limitation of the unsuspended AlN platform where the acoustic wave leaks into the substrate layer without contributing to the acousto-optic interaction. The platform is promising for studying interaction between cavity confined photons and propagating phonons of microwave frequency. In additional, the strong and high frequency acoustic waves realized in this platform can provide spatially-coherent time-domain modulation to induce non-reciprocity and break time-reversal symmetry in photonic systems[22,23]. At this high acoustic frequency, this system can also be applied for microwave photonics technology where optical and microwave channels of communication can be linked and interchanged[24].

**Acknowledgements**

We acknowledge the funding support provided by the Young Investigator Program (YIP) of AFOSR (Award Number FA9550-12-1-0338) and the National Science Foundation (Award Number ECCS-1307601). Parts of this work was carried out in the University of Minnesota Nanofabrication Center which receives partial support from NSF through NNIN program, and the Characterization Facility which is a member of the NSF-funded Materials Research Facilities Network via the MRSEC program.

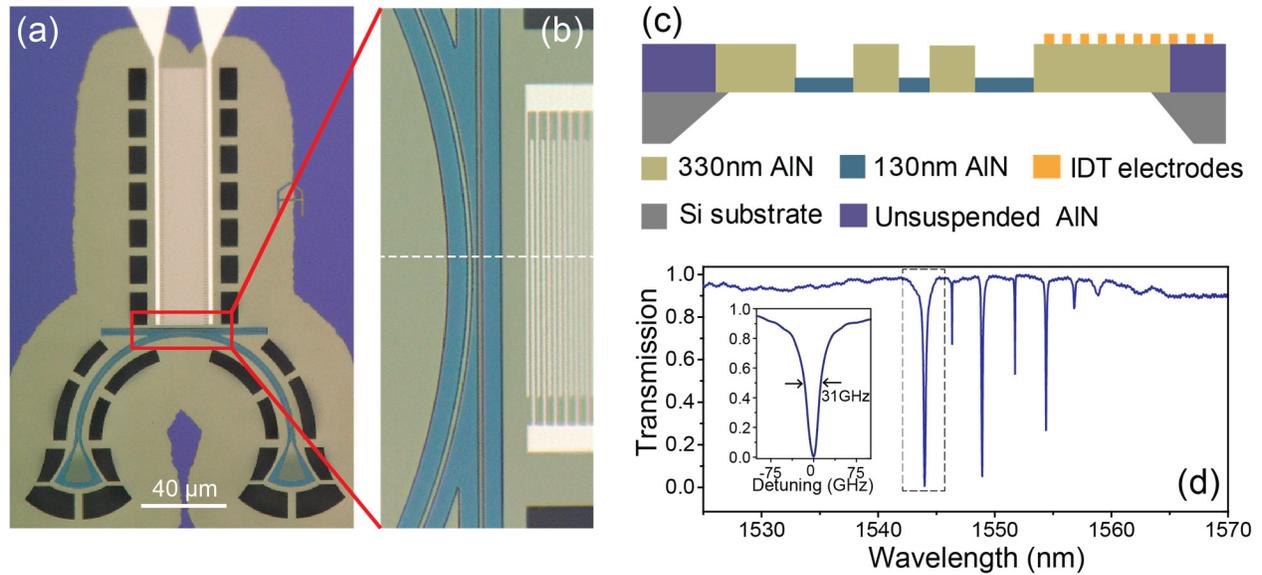

FIG. 1. (a) Optical microscope image of a fabricated device including IDT, photonic waveguide and nanocavity on a suspended AlN membrane (gray yellow regions). The unsuspended area of AlN is purple. (b) Zoomed in optical image of the area around the photonic crystal nanobeam cavity and IDT electrodes. The dark green region is the 130 nm AlN slab. (c) Cross-sectional view of the device structure along the white dashed line in (b). (d) Measured transmission spectrum of the photonic crystal nanocavity using a side-coupled waveguide. The inset displays a zoom-in view of the fundamental resonance mode, showing a linewidth of 31 GHz.

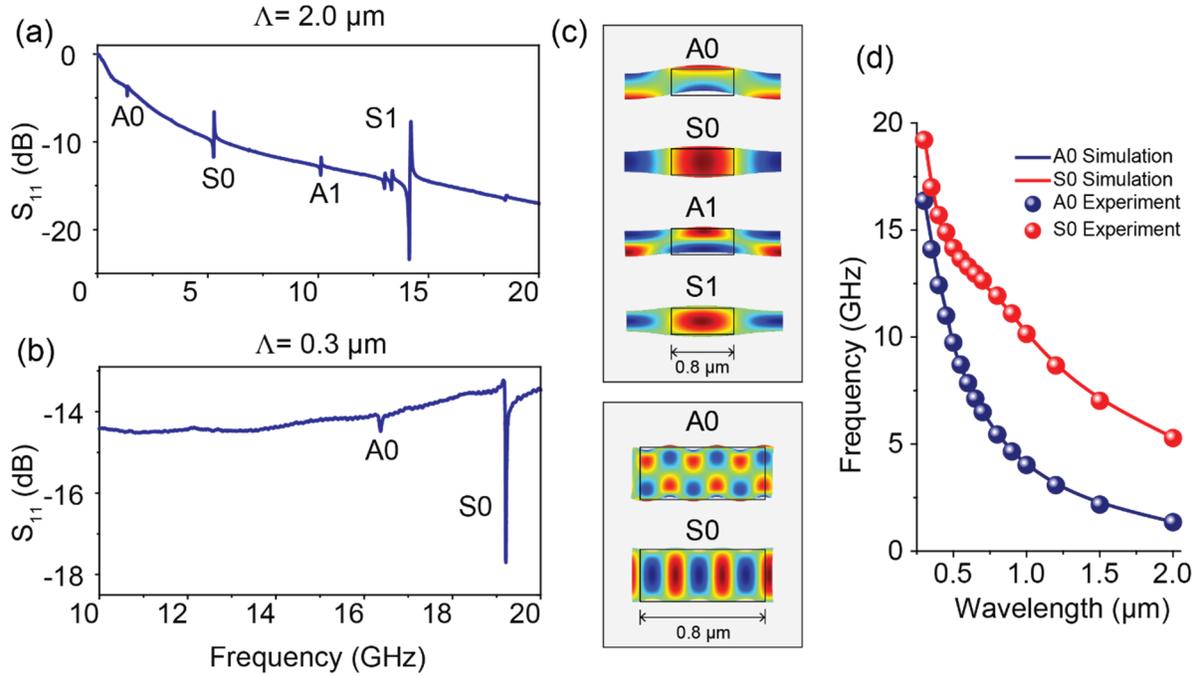

FIG. 2. (a) and (b) Measured spectra of the reflection coefficient $S_{11}$ of two devices with IDT periods of 2.0 μm and 0.3 μm, respectively. (c) Simulated Lamb wave modes for both devices. The deformation of the shape indicate the displacement and the color map illustrate the strain field. Red (blue) color indicates positive (negative) strain. The black box illustrate the cross-section of the nanocavity. "A" and "S" stand for "antisymmetric" and "symmetric". (d) The dispersion relation of Lamb wave in 330 nm thick AlN membrane. The solid lines are calculated results and the symbols depict experimentally measured values.

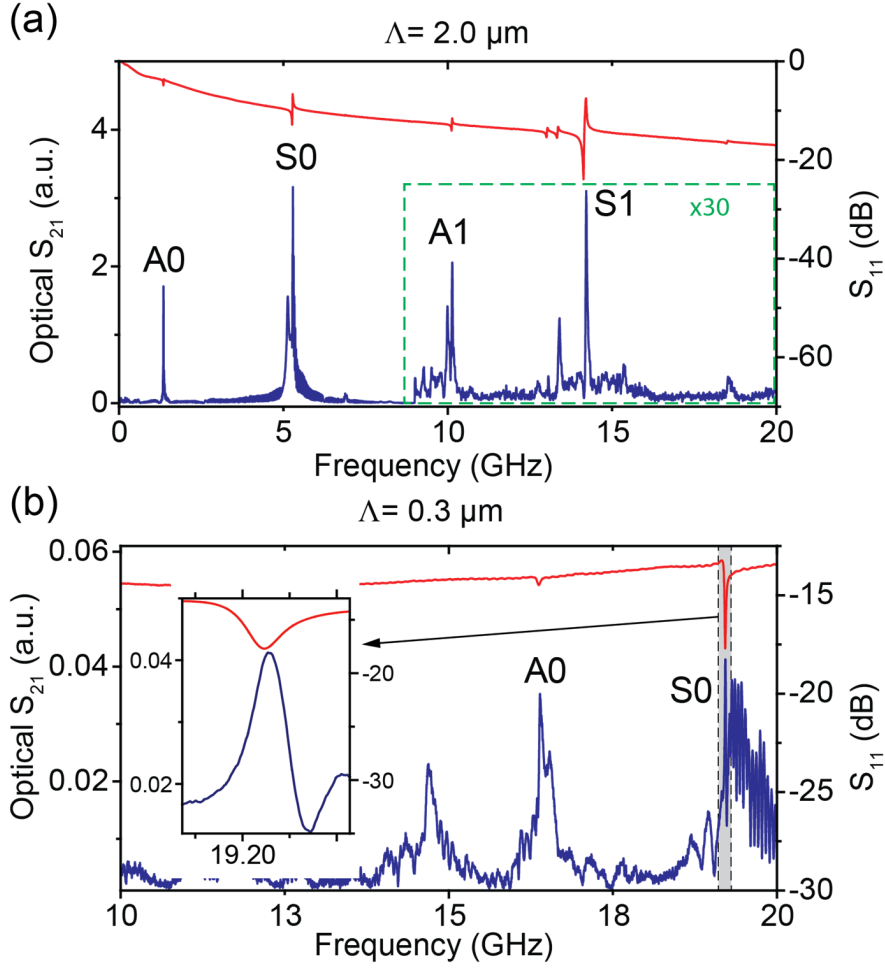

FIG. 3. Measured responses of the nanocavity to acousto-optic modulation induced by the Lamb wave, in terms of the spectra of the system's transmission coefficient $S_{21}$ (blue traces, left axes), for the devices with IDT period of, (a) 2.0 μm, and, (b) 0.3 μm. In (a), the spectrum in the dashed box is scaled up by 30 times for clarity. The spectra of $S_{11}$ (red traces, right axes) are also plotted for comparison. Inset of (b) is a zoom-in of the S0 mode at 19.20 GHz, of the device with 0.3 μm IDT.

**Table I. Measured acousto-optic modulation efficiency $G$.**

| Mode | $\Lambda = 2.0$ μm | | $\Lambda = 0.3$ μm | |
|---|---|---|---|---|
| | $f$ [GHz] | $G$ [MHz/mW$^{1/2}$] | $f$ [GHz] | $G$ [MHz/mW$^{1/2}$] |
| A0 | 1.35 | 760 | 16.39 | 6.35 |
| S0 | 5.28 | 1570 | 19.20 | 7.4 |
| A1 | 10.14 | 26 | | |
| S1 | 14.21 | 37 | | |

**Supplementary Online Material**

**Acousto-optic modulation of a photonic crystal nanocavity with Lamb waves in microwave K band**

Semere A. Tadesse[1,2], Huan Li[1], Qiyu Liu[1], and Mo Li[1]

*[1]Department of Electrical and Computer Engineering, University of Minnesota, Minneapolis, MN 55455, USA [2]School of Physics and Astronomy, University of Minnesota, Minneapolis, MN 55455, USA*

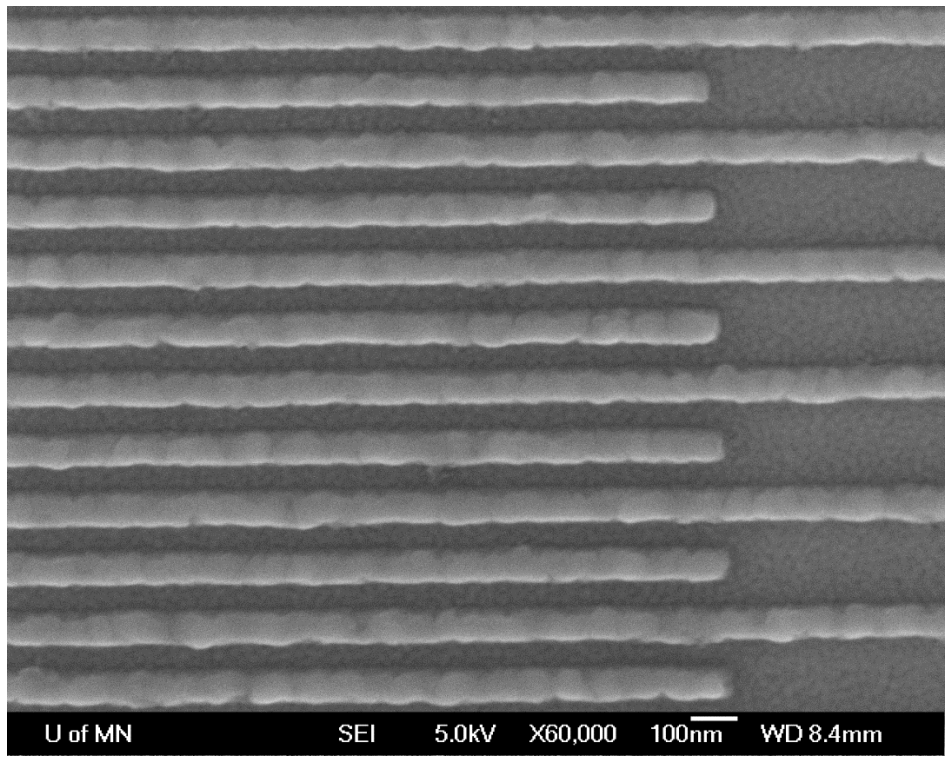

Figure S1 Scanning electron microscope image of the aluminum inter-digital transducer (IDT) with 0.3 μm period and 75 nm wide electrodes.